\documentclass[runningheads]{llncs}
\usepackage{graphicx, cite, pdfpages, colortbl, pifont}
\usepackage{multirow}
\usepackage{array}
\usepackage{hhline}
\usepackage{makecell}
\usepackage{floatpag}
\usepackage{booktabs}
\usepackage{hhline}
\usepackage{pgf}
\usepackage{tikz}
\usepackage{collcell}
\usepackage{adjustbox}
\usepackage{array}
\usepackage{multirow}

\begin{document}
\title{FetalNet: Multi-task Deep Learning Framework for Fetal Ultrasound Biometric Measurements}
%
%
\author{Szymon P\l otka\inst{1, 2, 4}, Tomasz W\l odarczyk\inst{2, 4}, Adam Klasa\inst{4}, Micha\l\space Lipa\inst{3}, Arkadiusz Sitek\inst{1}, Tomasz Trzci\'nski\inst{2, 5}}
\authorrunning{S. P\l otka et al.}
%
\institute{Sano Centre for Computational Medicine, Cracow, Poland \and Warsaw University of Technology, Warsaw, Poland \and Medical University of Warsaw, Warsaw, Poland \and Fetai Health Ltd. \and Tooploox, Wroclaw, Poland}
%
\maketitle              

\begin{abstract}

In this paper, we propose an end-to-end multi-task neural network called FetalNet with an attention mechanism and stacked module for spatio-temporal fetal ultrasound scan video analysis.
Fetal biometric measurement is a standard examination during pregnancy used for the fetus growth monitoring and estimation of gestational age and fetal weight. The main goal in fetal ultrasound scan video analysis is to find proper standard planes to measure the fetal head, abdomen and femur.
Due to natural high speckle noise and shadows in ultrasound data, medical expertise and sonographic experience are required to find the appropriate acquisition plane and perform accurate measurements of the fetus. In addition, existing computer-aided methods for fetal US biometric measurement address only one single image frame without considering temporal features. 
To address these shortcomings, we propose an end-to-end multi-task neural network for spatio-temporal ultrasound scan video analysis to simultaneously localize, classify and measure the fetal body parts. We propose a new encoder-decoder segmentation architecture that incorporates a classification branch. Additionally, we employ an attention mechanism with a stacked module to learn salient maps to suppress irrelevant US regions and efficient scan plane localization.
We trained on the fetal ultrasound video comes from routine examinations of 700 different patients. Our method called FetalNet outperforms existing state-of-the-art methods in both classification and segmentation in fetal ultrasound video recordings. The source code and pre-trained weights are publicly available \footnote[1]{https://github.com/SanoScience/FetalNet}.

\end{abstract}
\section{Introduction}
Fetal organ measurement using ultrasound (US) is currently the most popular way of assessing the state of the fetus’ growth and safety of the pregnancy \cite{liu2019deep}. It enables the operator to perform an array of measurements during single imaging session. Clinically, the most important are the measurements of biparietal diameter (BPD), head circumference (HC), femur length (FL) and abdominal circumference (AC). Fetal measurements enable obstetricians to evaluate fetal' growth and estimate the following parameters: gestational age (GA) and fetal weight (FW) \cite{hadlock1984estimating, hadlock1985estimation}.

In order to obtain proper fetal measurements, it is required to find a proper imaging plane (view) and follow strict guidelines that standardize the procedure. Both of those tasks require substantial knowledge and experience of the operator. Ultrasound images are characterized by high speckle noise, blur and shadows, which further increase the difficulty of the task. An automated way of measurement of the fetal body parts is meaningful because expert resources are scarce, especially in underdeveloped countries \cite{shah2015perceived, van2019automated}.
    
\textbf{Related Work:}
To automate the fetal body part measurements, researchers used computer-aided diagnosis, including the most advanced offered by deep learning. The problem of finding suitable views that meet the criteria of a standard measurement plane has been investigated \cite{baumgartner2017sononet, burgos2020evaluation}. Substantial research has been done to find the best algorithms that segment single fetal body parts. Encouraging results were achieved for segmentation of fetal heads \cite{sinclair2018human, zeng2021fetal} and abdomens \cite{ravishankar2016hybrid, sinclair2018cascaded}. Jang et al. \cite{jang2017automatic} proposed to use simultaneous segmentation and classification of abdomen, \cite{kim2018machine, budd2019confident} of head, and \cite{wu2017cascaded} both. 
Unlike ours, the large majority of algorithms developed to date focus on solving only one task at a time. 
Some address the task of choosing a good plane for fetal measurement, while others focus on segmenting a single fetal body part. 

To the best of our knowledge, the most similar method to ours is Liu et al. \cite{liu2020automated}. The authors developed a model that can classify and segment fetal head, abdomen and femur simultaneously. However, our method differs from \cite{liu2020automated} in the following aspects. First, their model was trained on single image frames and does not provide temporal fetal US scan video analysis. 
Second, their model always assigns one a body part to every  frame. This is an important limitation because ultrasound recordings contain many frames that do not contain any body part of interest appropriate for measurements and/or segmentation, and therefore their solution is not designed to work with ultrasound recordings. The paper \cite{prieto2021automated} has the capability of recognizing background (not appropriate for measurement frames). However, authors used the inpainting method to remove pixels with embedded annotations in retrospective cohort study data. Such image modification technique is not ideal in investigations that use deep learning and it is not practical.

In this paper, we propose an end-to-end pipeline called FetalNet that is designed to jointly localize, classify and measure the fetal body parts at the frame level. 
We examine the impact of temporal information extracted from frame sequence connected to the attention mechanism and stacked module on the fetal body parts measurement in the fetal US video recordings.

\textbf{Contributions:} The main contributions of our work are as follows: 
(i) we propose an end-to-end multi-task method called FetalNet for comprehensive $2D+t$ spatio-temporal fetal US video analysis to localize, classify and measure the fetal body parts simultaneously, (ii) we extend an attention gate mechanism by aggregating multi-scale feature maps of each decoder output to learn the local and global context of the fetal body structures that help to outperform both segmentation and classification state-of-the-art results.


\section{Method}
\label{sec:method}
\begin{figure}[ht!]
    \centering
    \includegraphics[width=12cm]{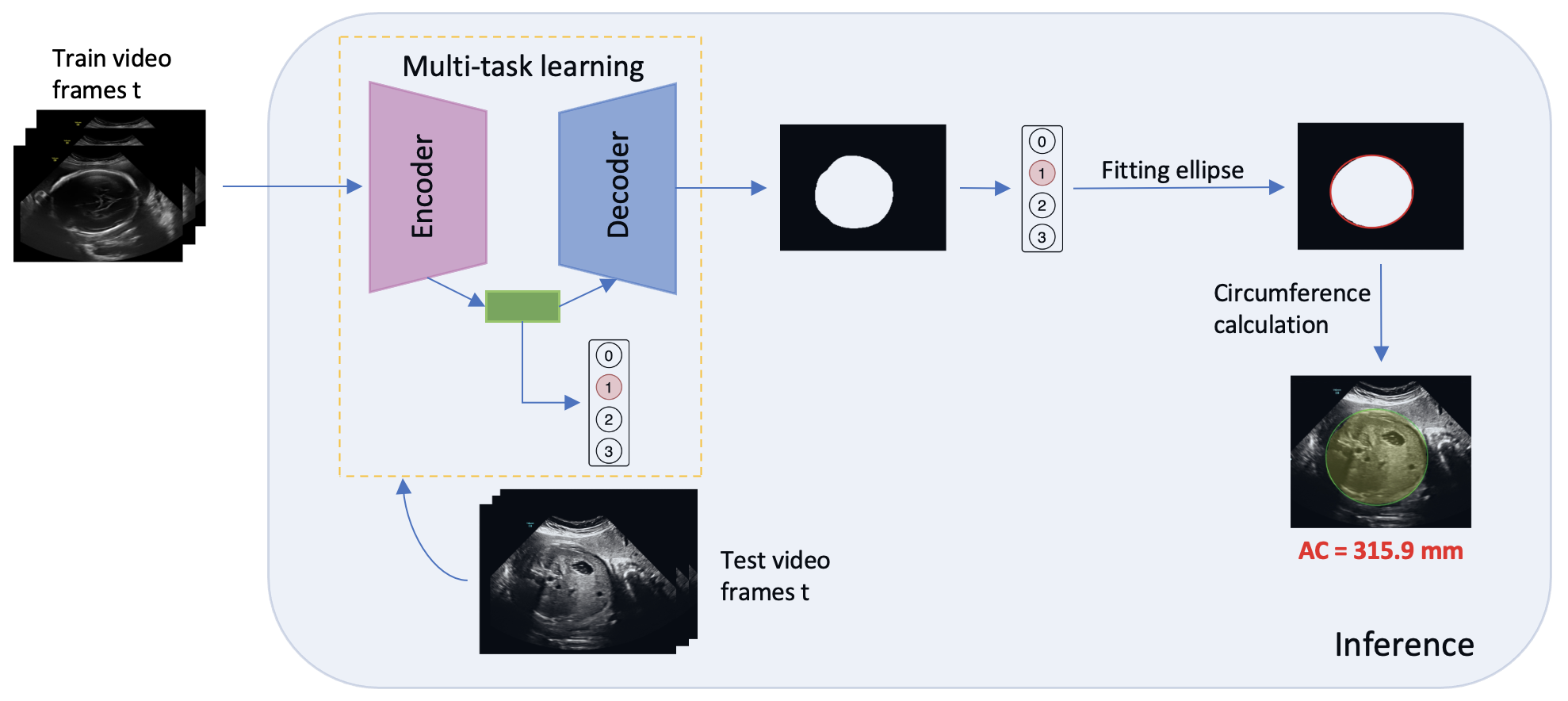}
    \caption{Overview of the proposed method. We train multi-task learning framework for simultaneous segmentation and classification using fetal US video sequences
    with four classes. Our encoder part extract high-level US features to 1) classify the fetal body parts and background, 2) to predict binary mask of fetal head, abdomen and femur. During inference, our method localizes and classifies each image at the frame level, and performs measurements respective to the predicted class.}
    \label{fig:method}
\end{figure}

In this section, we describe a multi-task learning neural network for spatio-temporal fetal US video analysis. Next, we describe how to automatically obtain measurements of each fetal body parts. Figure \ref{fig:method} shows an overview of our method called FetalNet for the automatic evaluation of fetal biometric measurement on fetal US scan video using a multi-task deep learning framework. 

\noindent

\begin{figure}[ht!]
    \centering
    \includegraphics[width=12.5cm]{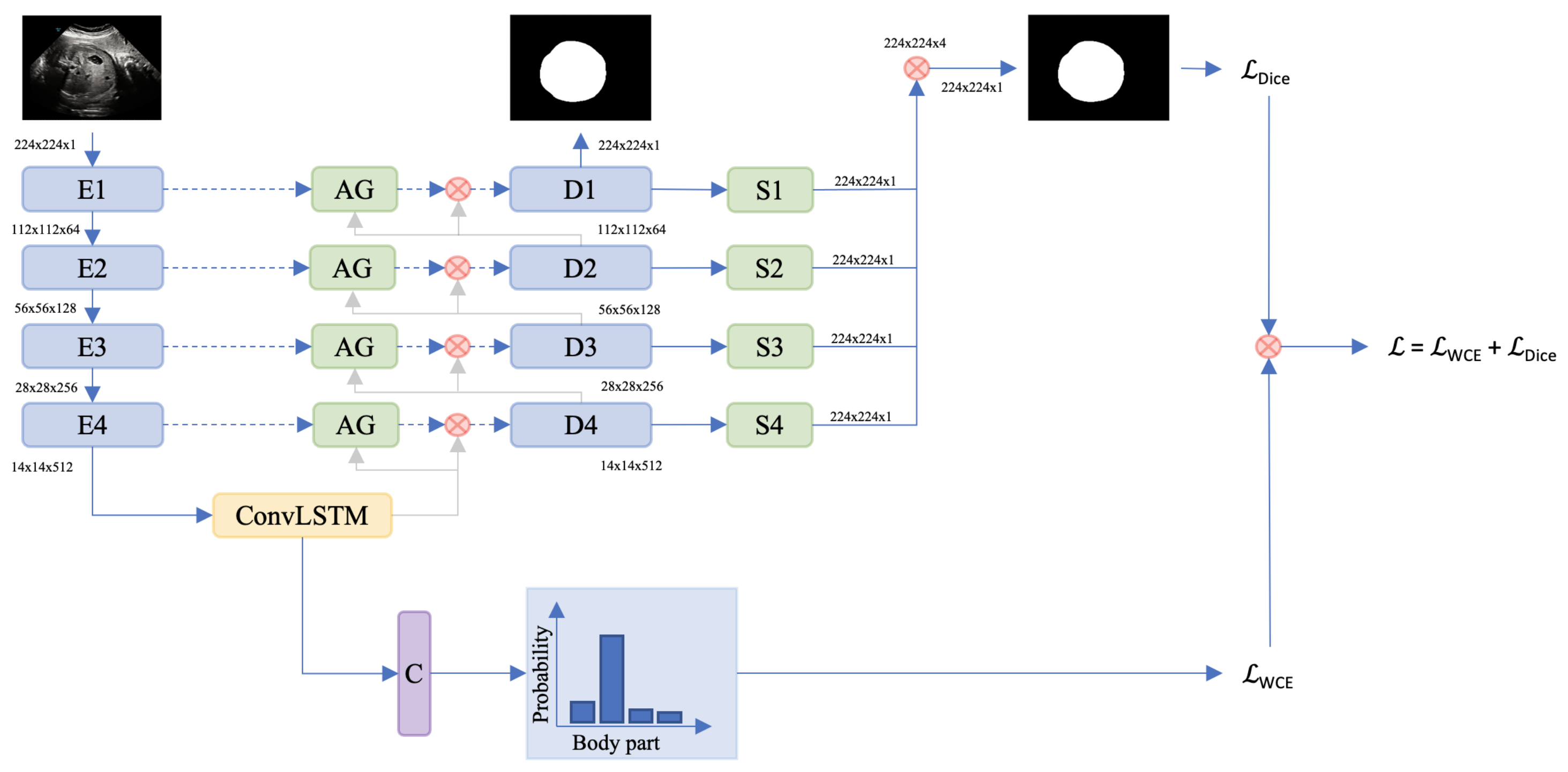}
    \caption{Overview of the proposed multi-task neural network. We base on an encoder-decoder U-Net neural network. As a bottleneck, we use the ConvLSTM module to learn image sequences. To focus on a region-of-interest (ROI) in the segmentation task, we concatenate encoders (E1-E4) by attention gates (AG) to decoders (D1-D4) via skip connections. To improve the performance of the segmentation and smooth our binary predicted masks, we use stacked multi-scale probability maps via up-sample convolutional outputs of the sides (S1-S4) probability maps to the input image size. For the classification task, we use ConvLSTM output features and a fully connected layer (C) with 4 outputs.}
    \label{fig:nnetwork}
\end{figure}
\noindent
\textbf{Multi-task neural network:} Inspired by \cite{mehta2018net, wang2018simultaneous}, we use an encoder-decoder based convolutional neural network (CNN) for simultaneous segmentation and classification of the fetal body parts on the fetal US video sequences. We use an encoder part to extract high-level US image features. The output of an encoder is forward to the ConvLSTM-based bottleneck. The ConvLSTM cells are able to retain spatial-temporal US image features in memory, which can effectively improve the performance precision and accuracy of both classification and segmentation. Due to various shapes and sizes in our dataset, we employ an attention gate mechanism to implicitly learn to suppress irrelevant regions in an input video sequence, while highlighting the salient features of the target region-of-interest. Attention gate mechanism helps to better exploit local information to efficient object localize (i.e. fetal body parts) and improve prediction performance. Every encoder block forwards its output feature maps and concatenates them with an attention gate to the decoder part. In fetal examinations, fetal body parts are hardly visible, and the sonographer's manual examination relies heavily upon low-level semantic information to draw boundaries. To improve the performance of binary prediction feature maps, we employ deep supervision to connect multi-scale lower and higher-level of each decoder features together \cite{harrison2017progressive} called stacked module. Multi-scale feature maps help to encode both global and local context. We use a set of $3 \times 3$ 2D convolutional layers to up-sample the feature maps after each convolutional block. Then, we combine the previous high-level feature maps to aggregated binary segmentation map. As our ablation study demonstrates, a stacked module with an attention gate can significant impact segmentation and reduce the measurement error of the fetal body parts over both attention U-Net and stacked module U-Net. For the classification branch, we use spatial-temporal ConvLSTM-output features to classify each of the following classes: fetal head, abdomen, femur or background at the frame level. Figure \ref{fig:nnetwork} shows the proposed multi-task learning method called FetalNet for spatio-temporal fetal ultrasound scan video analysis.
\newline
\noindent
\newline
\textbf{Biometric measurement:} First, we resize the segmentation output of our multi-task neural network, which comes in the form of a binary mask to the size of the input image. Next, we apply binary thresholding $p = 0.6$ and perform erosion followed by dilation, using a $5 \times 5$ cross-shaped structuring element. This ensures that the predicted masks are sufficiently denoised. Finally, we use a median blur filter to smooth the edges of the predicted masks. Depending on the classification results, we use different methods to obtain adequate measurements. For HC and AC, firstly, we use a function to find contours of segmentation output. Next, we use Ramer-Douglas-Peucker approximation \cite{ramer1972iterative, douglas1973algorithms} to enhance the accuracy of the subsequent ellipse fitting, for which perimeter is calculated and stored. Additionally, to acquire the measurement of BPD, we store the length of the short axis of the ellipse fitted to the head. To obtain FL, we precisely fit a rectangular bounding box to the contours of the predicted binary mask found by using the same function as in case of head and abdomen. Next, we store the length of the fitted rectangle. Finally, we convert all of the obtained pixel-valued measurements to centimetres by multiplying them by pixel spacing, an attribute that encodes the physical distance between centres of the pixels, stored in DICOM files metadata.

\section{Experiments and Results}
In this section, we introduce a novel Fetal dataset and show the performance of our method called FetalNet on this dataset.
\noindent
\newline
\textbf{Fetal dataset:} Our dataset consists of 700 two-dimensional (2D) fetal US video sequences examinations of head, abdomen and femur and comes from 700 different patients. Each 2D fetal US video sequence consists of between 250 and 460 frames. Overall our dataset consists of over 274,000 frames. The data comes from volunteer pregnant women with pregnancies between 15th and 38th weeks of gestation, acquired during a routine clinical screening examination.  Data was acquired using five different GE Voluson ultrasound scanners (E8, E10, S6, S8, P8) at two different resolution: 975 $\times$ 742 pixels or 1100 $\times$ 960 pixels.  From video sequences sonographers identified standard views that are suitable to perform the measurement and annotated them. Overall, our annotated dataset consists of 62324 standard views and 211951 background views. We used the number of examples for each class for training and validation: 32215 heads, 26403 abdomens, 3706 femurs and 211951 backgrounds. The background class shows indistinguishable structures around standard view plane frames. 

\begin{figure}
   \centering
\begin{tabular}{ccc}
\includegraphics[width=4cm]{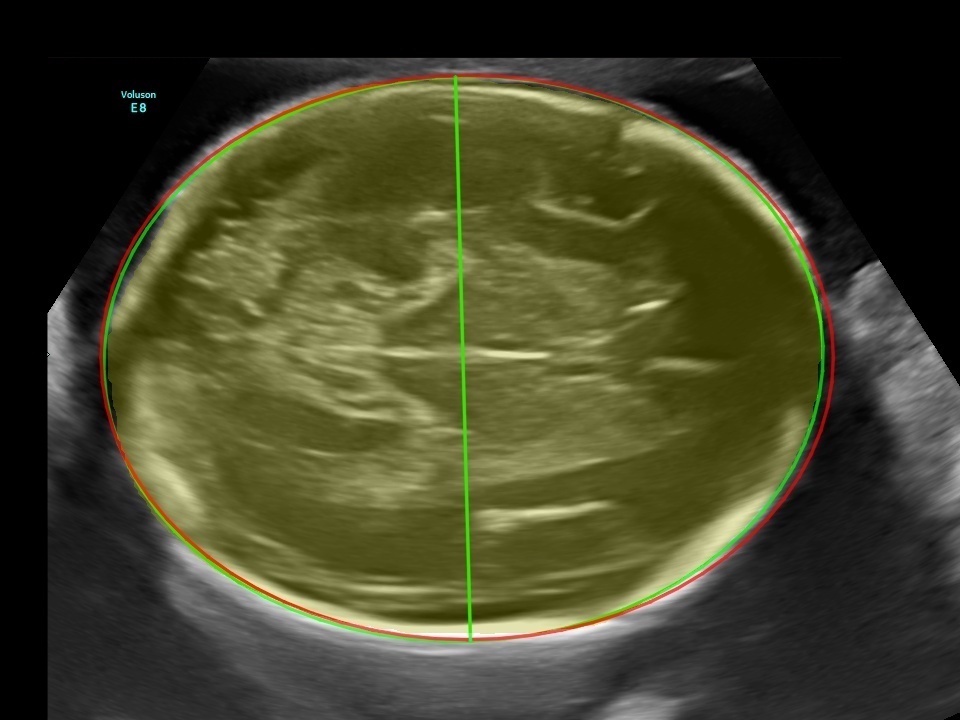}&
\includegraphics[width=4cm]{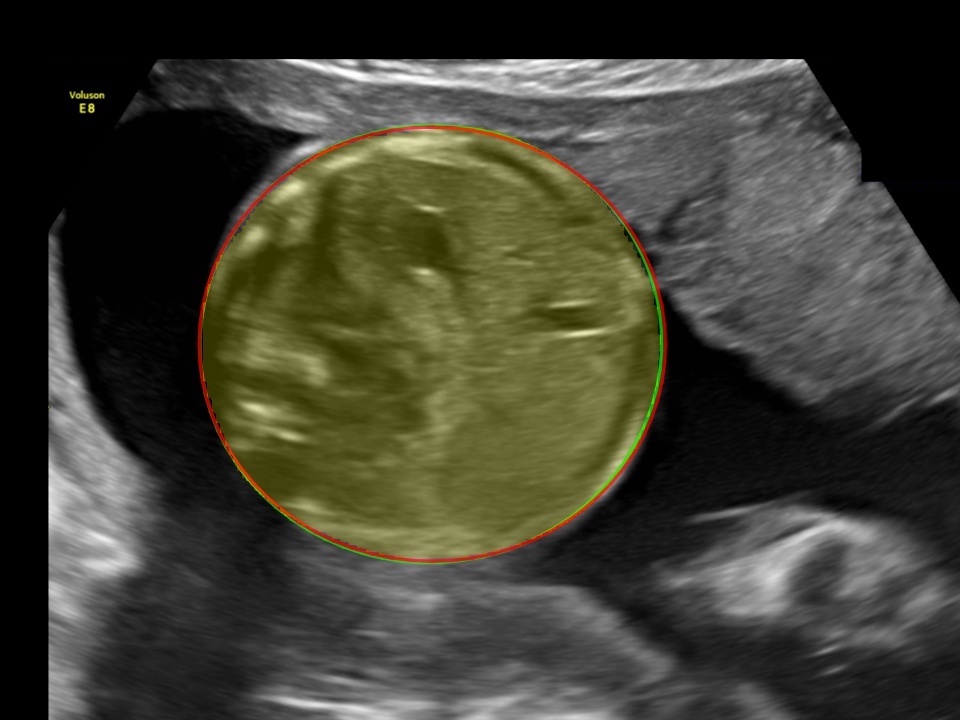}&
\includegraphics[width=4cm]{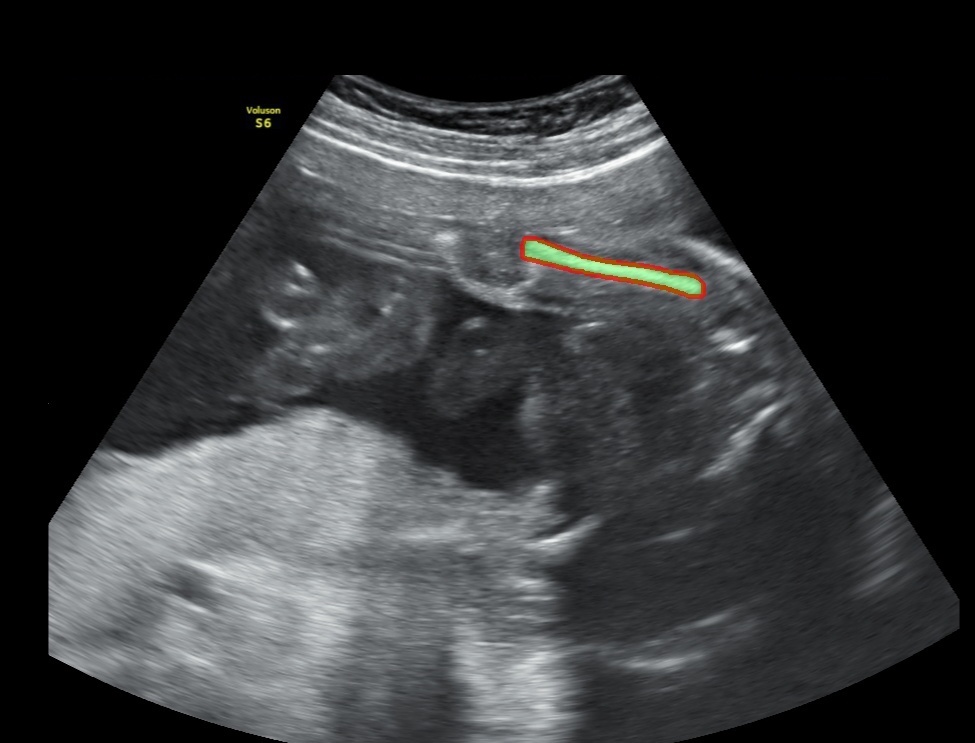} \\
\includegraphics[width=4cm]{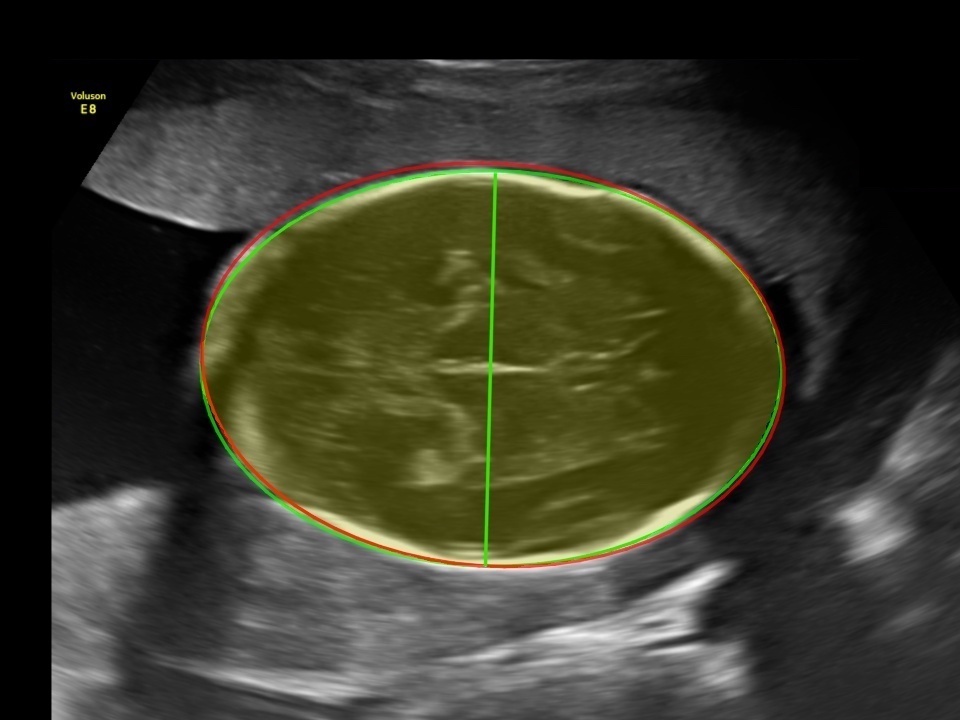}&
\includegraphics[width=4cm]{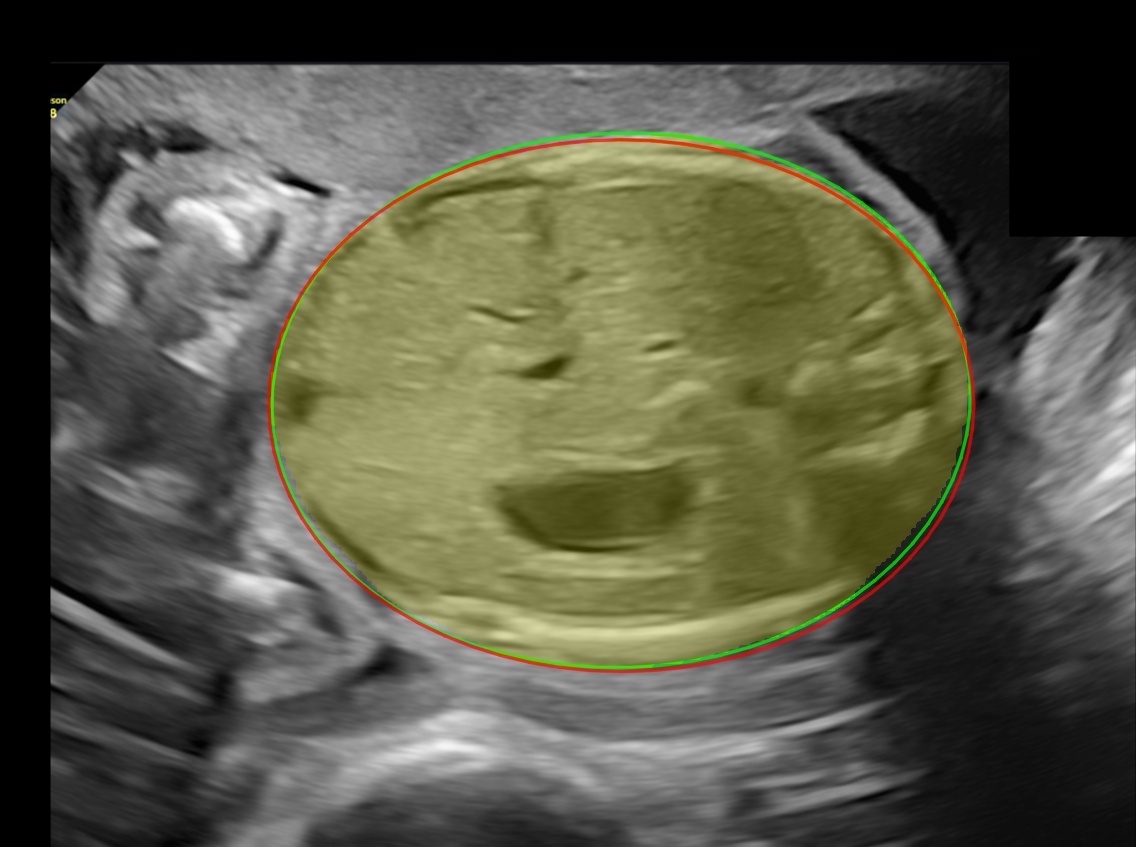}&
\includegraphics[width=4cm]{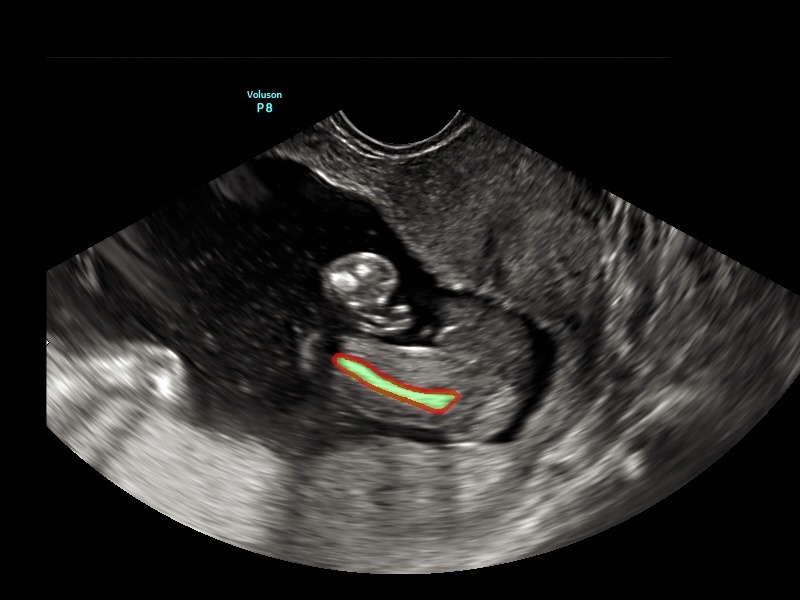} \\
\includegraphics[width=4cm]{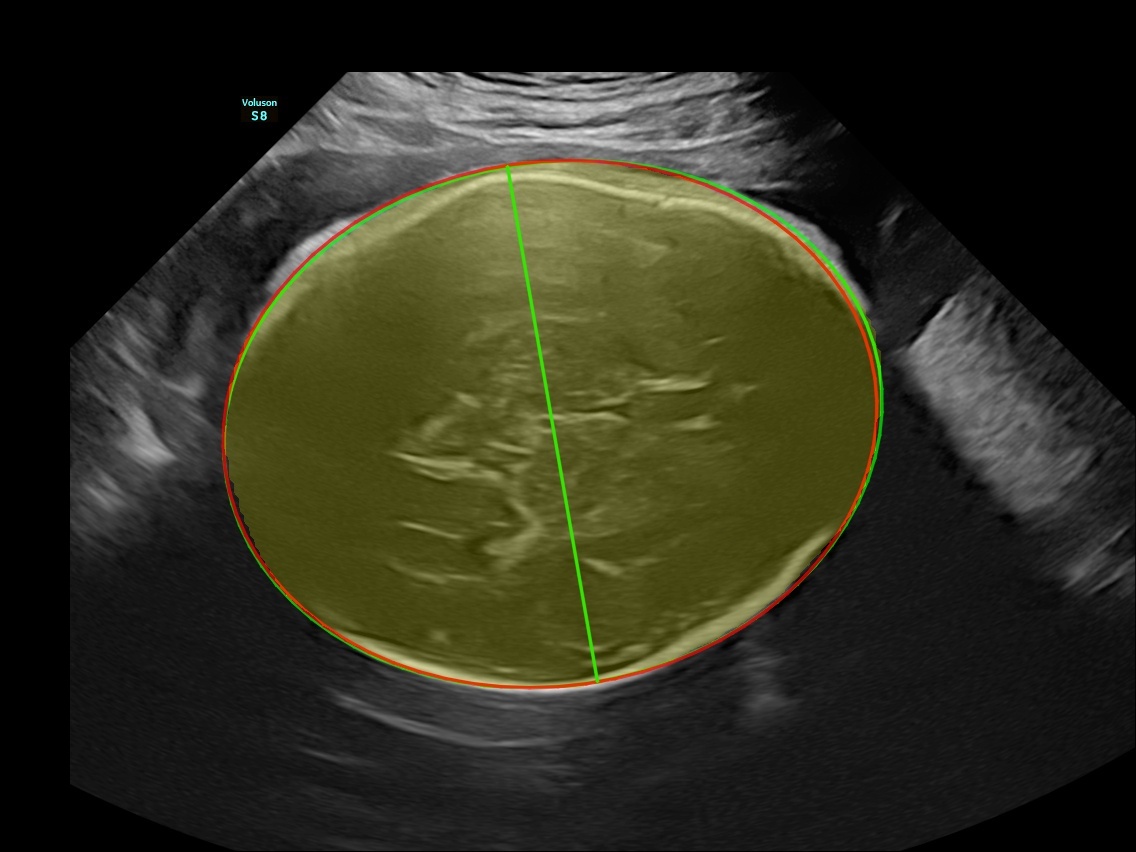}&
\includegraphics[width=4cm]{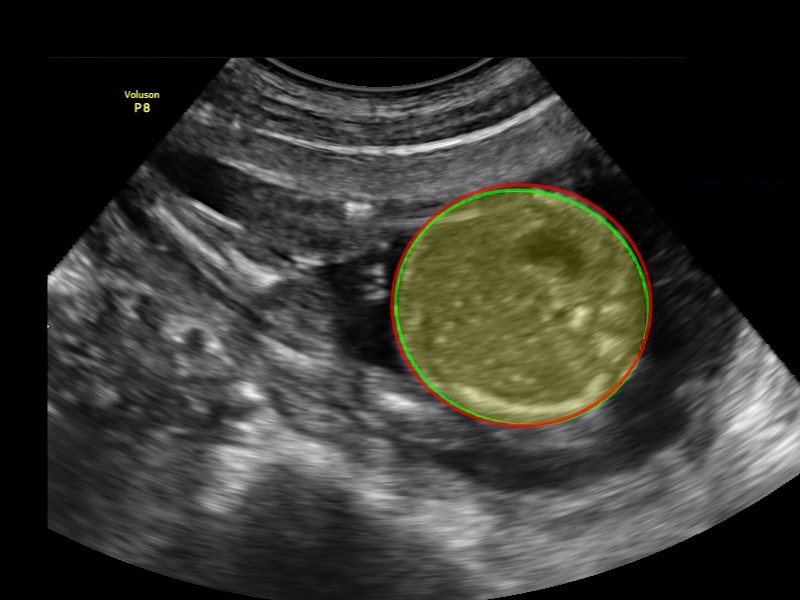}&
\includegraphics[width=4cm]{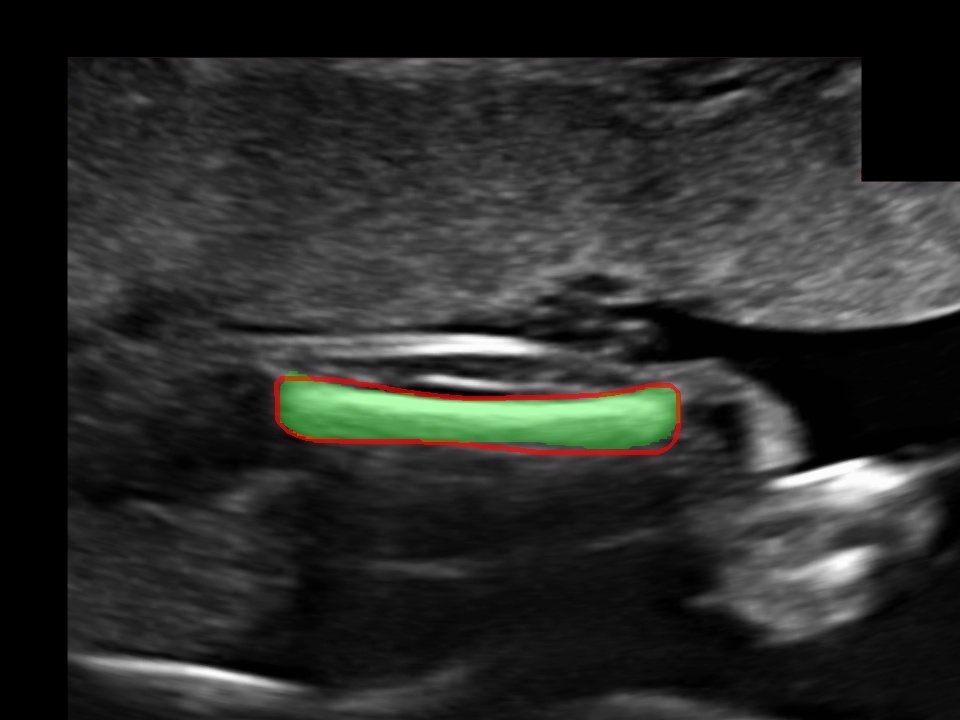} \\
\includegraphics[width=4cm]{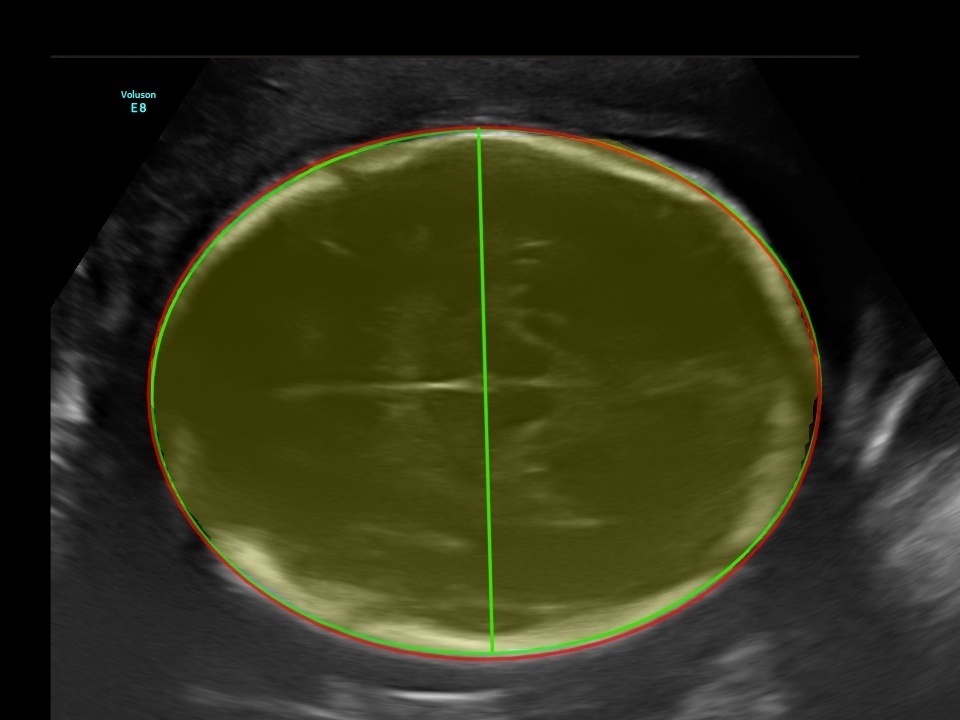}&
\includegraphics[width=4cm]{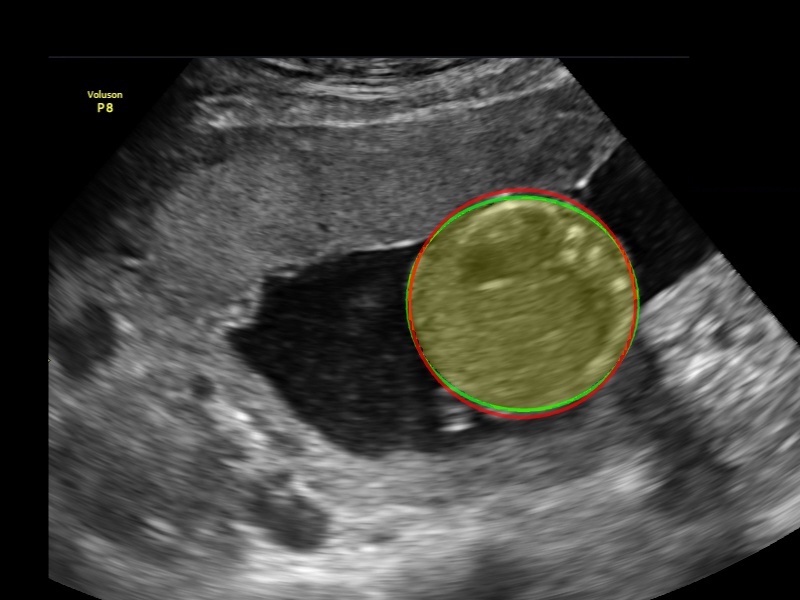}&
\includegraphics[width=4cm]{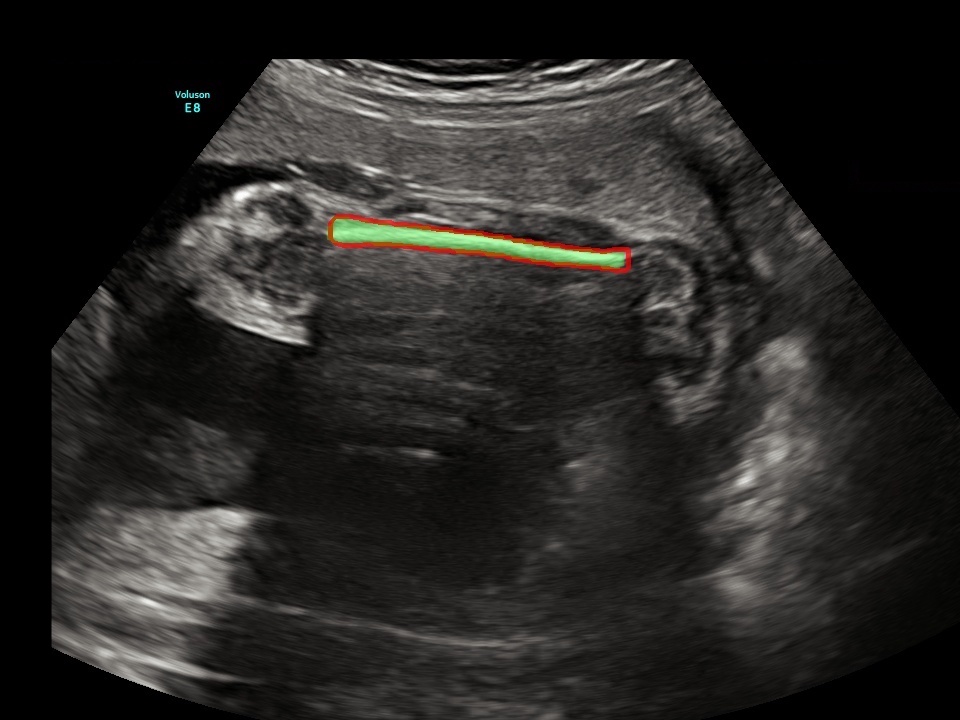} \\
\includegraphics[width=4cm]{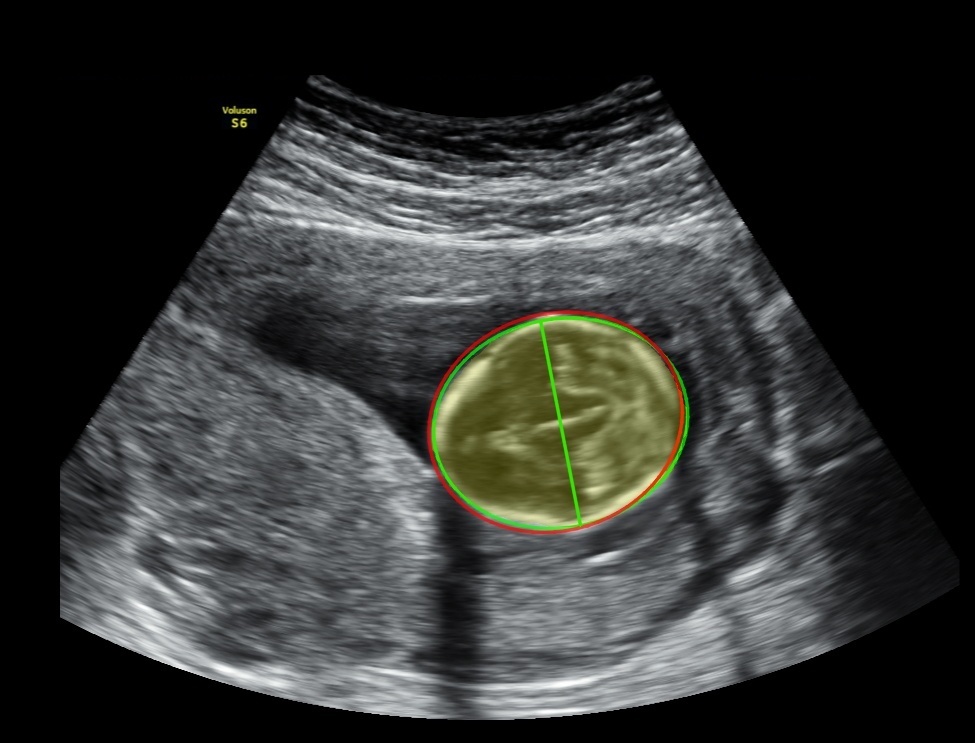}&
\includegraphics[width=4cm]{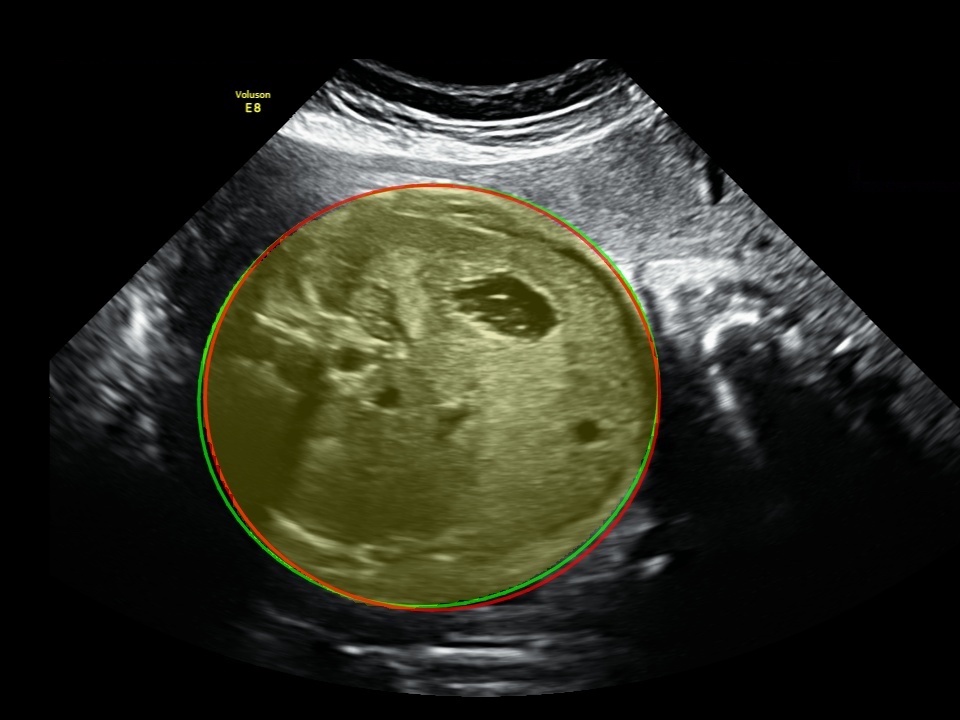}&
\includegraphics[width=4cm]{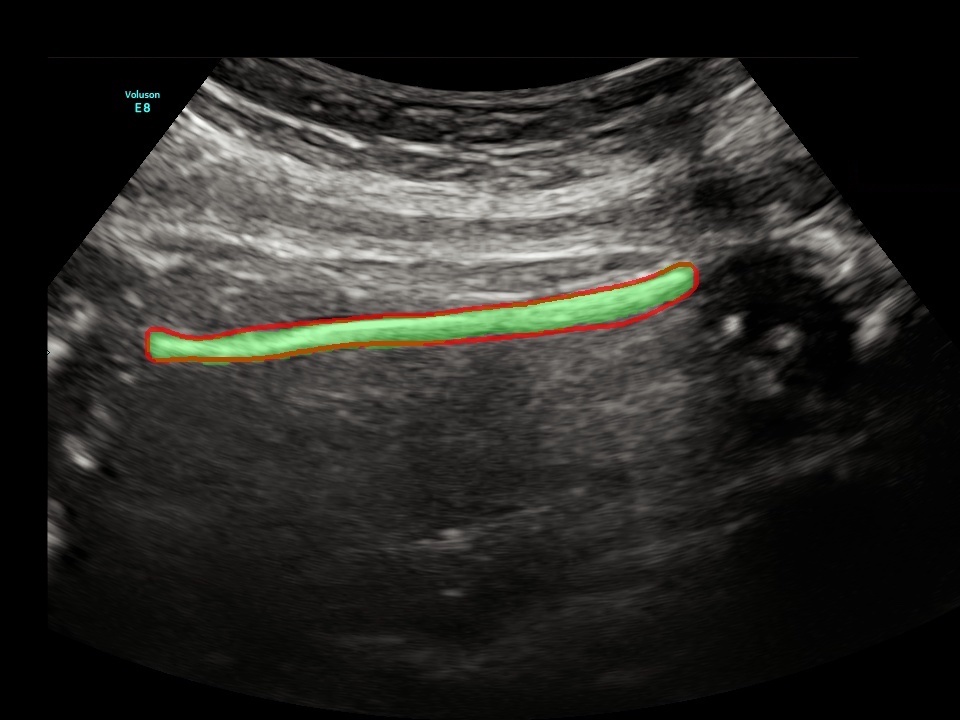} \\
a) head & b) abdomen & c) femur\\
\end{tabular}

    \caption{We show results on our test set. We show five examples per class on test data from patients between 15 and 38 weeks of gestation. From the left to right: a) HC and BPD b) AC and c) FL measurement obtained by our multi-task deep learning framework. The red line represents a ground truth, while the green line our predictions. As shown, our method can classify and measure with good precision fetal body parts with different size and data quality.}
    \label{fig:seg-results}
\end{figure}

The data come from six different research institutes and were anonymized before use in this study. Six sonographers with experience (40, 25, 20, 20, 15, 8 years, respectively) provided ground truths for the dataset in the form of annotations drawn on the anonymized images and values of the performed measurements. The annotations of heads and abdomens were ellipses similar to those that were used for manual measurements typically done at the ultrasound scanner. Annotations of femurs were created by free-hand drawings of their outlines.
\newline
\textbf{Preprocessing:} First of all, we transform 2D fetal US video sequences into separate ordered frames. Then, we anonymize raw 2D US images by removing personal data displayed on the top of the images. Next, we remove unnecessary text burned in images like device settings. In the next step, we convert raw DICOM data into PNG files. Finally, we randomly split our dataset into 60\% training (420 cases), 20\% validation (140 cases) and 20\% test (140 cases) set.
\newline
\textbf{Data Augmentation:} 
During training, to prevent overfitting and make the neural network more robust, we apply various data augmentation techniques: random (i) rotation between -15 and 15 degrees, (ii) brightness, (iii) contrast, (iv) horizontal and (v) vertical flip. We also use a shuffled sampler.
\newline
\textbf{Evaluation Metrics:} We use the Jaccard Index (IoU) and Dice Similarity Coefficient (DSC) for the segmentation task as evaluation metrics. For classification, we use accuracy, precision, recall and F1-score. Finally, we evaluate measurement performance using Absolute Difference (ADF).
\newline
\textbf{Implementation details:} We base our network on U-Net \cite{ronneberger2015u}, our encoder-decoder includes eight convolutional blocks, four in the encoder part and four in the decoder part. We concatenate encoders with attention gates \cite{oktay2018attention} to decoders via skip-connections. We use an attention gate mechanism to focus on certain parts of images. To improve the performance of the segmentation and smooth our binary predicted masks, we use stacked probability multi-scale feature maps via up-sample convolutional outputs of the side feature maps to the input image size. We extend the original U-Net implementation that each block consists of the following order: Conv3x3-BatchNorm-ReLU-Conv3x3-BatchNorm-ReLU-Dropout2D with $p=0.2$. After each block of the encoder part, we apply Max Pooling layer with a kernel size of $2 \times 2$ and stride = 2. The number of feature maps in the input layer is equal to n = 64. The rest of the eight convolutional blocks consists of 2n-4n-8n-16n-8n-4n-2n-n feature maps. For the classification branch, we apply Adaptive Average Pooling 2D and Dropout2D with $p=0.4$ as ConvLSTM output before Fully Connected layer with $14 \times 14 \times 16n$ feature maps on the output.
We use Adam as an optimizer with a learning rate of $10^{-4}$, weight decay of $10^{-5}$ and batch size of 16 for 80 epochs. As loss function, we use the sum of Dice Loss $\mathcal{L}_{dice}$ and Weighted Cross-Entropy Loss $\mathcal{L}_{WCE}$ with the following weights per class 0.25, 0.25, 0.4 and 0.1, respectively: 
\begin{equation}
    \mathcal{L} = \mathcal{L}_{dice} + \mathcal{L}_{WCE}
\end{equation}
Our training set contains an overall of 87771 fetal US images and annotations of: 26072 fetal heads, 20901 abdomens, 2956 femurs and 169500 of background as 0, 1, 2, 3 class, respectively. We scale all images and annotation masks to 224 $\times$ 224 pixels. 
We train a neural network on NVIDIA Titan RTX 24GB GPU for 240h.
We implement our neural network in Python based on PyTorch deep learning library.
Figure \ref{fig:nnetwork} shows proposed neural network.
\newline
\textbf{Segmentation results:}
We evaluate our model on 57001 test images of fetal head (7250 images), abdomen (6580 images), femur (720 images) and background (42451 images) class. For segmentation, we use Jaccard Index, also known as Intersection over Union (IoU) and the Dice similarity coefficient (DSC) as the evaluation metrics. We obtain the following results: 0.905 and 0.962 for IoU and DSC, respectively.
The qualitative results of our proposed network are depicted in Figure \ref{fig:seg-results}. It can be seen that our method was able to localise fetal head, abdomen and femur, which are subject to variability in scale and appearance.
\newline
\textbf{Classification results:} For classification as the evaluation metrics, we use accuracy, precision, recall and F1 score. We obtained the following results: 0.96, 0.97 and 0.96 for precision, recall and F1 score, respectively. Table ~\ref{tab:segclsresults} demonstrates FetalNet classification results against base U-Net, Deeplabv3 \cite{chen2017rethinking}, FCN-8s and FCN-32s \cite{long2015fully}. The proposed system outperforms the state-of-the-art neural networks in terms of head, abdomen and femur in segmentation and classification accuracy.
\begin{table}[]
\centering
\caption{Comparison of segmentation and classification of fetal body parts: head, abdomen, femur and background with the state-of-the-art neural networks and FetalNet}
\label{tab:segclsresults}
{%
\begin{tabular}{l|c|c|c|c|c|c}
\textbf{Method} & \multicolumn{1}{l|}{\textbf{IoU}} & \multicolumn{1}{l|}{\textbf{Dice}} & \textbf{Accuracy} & \textbf{Precision} & \textbf{Recall} & \textbf{F1}  \\ \hline
U-Net (base) & 0.862  & 0.921 & - & - & - & - \\
DeepLabv3 & 0.851 & 0.912 & 0.922 & 0.91 & 0.89 & 0.90  \\
FCN-8s & 0.865 & 0.924 & 0.933 & 0.93 & 0.91 & 0.92  \\
FCN-32s & 0.872 & 0.932 & 0.935 & 0.93 & 0.91 & 0.92  \\
\textbf{FetalNet (ours)} & \textbf{0.905} & \textbf{0.962} & \textbf{0.975} & \textbf{0.96} & \textbf{0.97} & \textbf{0.96}
\end{tabular}
}
\end{table}

Our method outperforms the current state-of-the-art methods in both classification and segmentation (Table \ref{tab:segclsresults}), and the fetal body part measurement (Table \ref{tab:measurementresults}). As shown in Figure \ref{fig:seg-results}, our segmentation results of standard view scans are comparable to the ground-truths provided by experienced sonographers.

\noindent
\textbf{Measurement results:} Table \ref{tab:measurementresults} reports results of fetal head, abdomen and femur error measurement (in mm) against state-of-the-art neural networks measured as the mean and standard deviation.

\begin{table}[]
\centering
\caption{Comparison of measurement error (in mm) of fetal body parts: head, abdomen, femur with the state-of-the-art neural networks and FetalNet.}
\label{tab:measurementresults}
{%
\begin{tabular}{l|c|c|c}
\textbf{Method} & \multicolumn{1}{l|}{\textbf{Head}} & \multicolumn{1}{l|}{\textbf{Abdomen}} & \textbf{Femur}  \\ \hline
U-Net (base) & 4.5±3.2  & 5.4±3.1 & 1.5±1.4 \\
DeepLabv3 & 4.8±3.4 & 5.5±3.4 & 1.5±1.3  \\
FCN-8s & 4.7±3.1 & 5.3±3.3 & 1.6±1.2  \\
FCN-32s & 3.9±2.8 & 4.9±3.2 & 1.2±0.8  \\
\textbf{FetalNet (ours)} & \textbf{2.9±1.2} & \textbf{3.8±3.0} & \textbf{0.8±1.2}
\end{tabular}
}
\end{table}

\noindent
\textbf{Ablation study:} We conduct the ablation study to show the effectiveness of the proposed method called FetalNet in terms of both segmentation and classification. We use the same dataset and hyperparameters of the neural network for each experiment, if not mentioned. Table \ref{tab:ablation_study} shows the experiments for the proposed method with different combinations of modules. As we can see, multi-task learning improves segmentation results to compare with U-Net base model. For the segmentation, combining the proposed $\mathcal{L}_{WCE}$ performed better than only using Dice loss. The results were further improved after introducing attention gate mechanism and stacked modules separately. Finally, our model trained with each of the proposed extensions, resulting in a notable performance gain over all the metrics.

\begin{table}[]
\centering
\caption{Evaluation of different components of neural network on our dataset. Results show that the combination of all components gives the best performance.}
\label{tab:ablation_study}
\resizebox{\textwidth}{!}{%
\begin{tabular}{l|c|c|l|l|c}
\textbf{Method} & \multicolumn{1}{l|}{\textbf{Segmentation}} & \multicolumn{1}{l|}{\textbf{Classification}} & \textbf{IoU} & \textbf{Dice} & \multicolumn{1}{c}{\textbf{Accuracy}} \\ \hline
U-Net (base) & \ding{51}  & - & 0.862 & 0.921 & - \\
U-Net+cls & \ding{51} & \ding{51} & 0.872 & 0.945 & 0.961 \\
U-Net+cls+AG & \ding{51} & \ding{51} & 0.891 & 0.951 & 0.971 \\
U-Net+cls+SM & \ding{51} & \ding{51} & 0.893 & 0.954 & 0.972 \\
\textbf{U-Net+cls+AG+SM (ours)} & \ding{51} & \ding{51} & \textbf{0.905} & \textbf{0.962} & \textbf{0.975}
\end{tabular}
}
\end{table}

In this paper, we propose an end-to-end multi-task method called FetalNet for spatio-temporal full-length routine fetal US scan video analysis. In particular, we consider attention mechanism and present how to incorporate it into fetal biometric measurement to better exploit local structures. We introduce aggregation of multi-scale feature maps as a stacked module making our approach more robust to spatio-temporal fetal US scan video analysis, where the previous methods fail. This allows for accurate and precise simultaneous localization and classification of the fetal body parts in freehand fetal US video recordings. Our method incorporates a classification branch to the U-Net-based encoder-decoder neural network. To learn temporal features, we employ the ConvLSTM layer as a bottleneck. To make our method more robust on ultrasound noise and shadow, we exploit an attention gate mechanism to focus on relevant ROIs at the frame level. We introduce a stacked module, aggregating the multi-scale feature maps of the decoder to learn the local and global context of the target. The ablation study (Table \ref{tab:ablation_study}) shows that using both additional modules, our methods achieve better results in segmentation and classification of the fetal body parts.

\section{Conclusions}

In this paper, we proposed an end-to-end multi-task method called FetalNet for spatio-temporal fetal US scan video analysis. FetalNet is designed to jointly localize, classify and measure the fetal body parts during routine freehand fetal US examinations. The proposed method has the potential as fetal biometry assistance tool for clinical use by non-experienced personnel. Usage of our approach in a clinical environment requires real-time feedback for a sonographer during routine fetal US examinations. Due to the large size of model parameters, we will implement a more efficient neural network to work on computationally low-cost devices. Additionally, we will improve and make our model more robust by adding to the training set data generated on a larger variety of devices as well as low-quality data to simulate data acquisitions made by non-expert personnel. We will also extend our method to automatically detect abnormalities of the fetus and perform a direct estimation of gestational age and fetal weight.

\section*{Acknowledgements}

The authors would like to thank the following medical sonographers for data, annotations and clinical expertise: Jan Klasa, MD; Bogusław Marinković, MD; Wojciech Górczewski, MD; Norbert Majewski, MD; Anita Smal-Obarska, MD. This research is supported by the European Union’s Horizon 2020 research and innovation programme under grant agreement Sano No 857533 and the International Research Agendas programme of the Foundation for Polish Science, co-financed by the European Union under the European Regional Development Fund and by Warsaw University of Technology (grant of the Scientific Discipline of Computer Science and Telecommunications agreement of 18/06/2020).
%
%
%
\bibliographystyle{splncs04}
\bibliography{references}

\begin{thebibliography}{10}
\providecommand{\url}[1]{\texttt{#1}}
\providecommand{\urlprefix}{URL }
\providecommand{\doi}[1]{https://doi.org/#1}

\bibitem{baumgartner2017sononet}
Baumgartner, C.F., Kamnitsas, K., Matthew, J., Fletcher, T.P., Smith, S., Koch,
  L.M., Kainz, B., Rueckert, D.: Sononet: real-time detection and localisation
  of fetal standard scan planes in freehand ultrasound. IEEE transactions on
  medical imaging  \textbf{36}(11),  2204--2215 (2017)

\bibitem{budd2019confident}
Budd, S., Sinclair, M., Khanal, B., Matthew, J., Lloyd, D., Gomez, A.,
  Toussaint, N., Robinson, E.C., Kainz, B.: Confident head circumference
  measurement from ultrasound with real-time feedback for sonographers. In:
  International Conference on Medical Image Computing and Computer-Assisted
  Intervention. pp. 683--691. Springer (2019)

\bibitem{burgos2020evaluation}
Burgos-Artizzu, X.P., Coronado-Guti{\'e}rrez, D., Valenzuela-Alcaraz, B.,
  Bonet-Carne, E., Eixarch, E., Crispi, F., Gratac{\'o}s, E.: Evaluation of
  deep convolutional neural networks for automatic classification of common
  maternal fetal ultrasound planes. Scientific Reports  \textbf{10}(1),  1--12
  (2020)

\bibitem{chen2017rethinking}
Chen, L.C., Papandreou, G., Schroff, F., Adam, H.: Rethinking atrous
  convolution for semantic image segmentation. arXiv preprint arXiv:1706.05587
  (2017)

\bibitem{douglas1973algorithms}
Douglas, D.H., Peucker, T.K.: Algorithms for the reduction of the number of
  points required to represent a digitized line or its caricature.
  Cartographica: the international journal for geographic information and
  geovisualization  \textbf{10}(2),  112--122 (1973)

\bibitem{hadlock1984estimating}
Hadlock, F.P., Deter, R.L., Harrist, R.B., Park, S.: Estimating fetal age:
  computer-assisted analysis of multiple fetal growth parameters. Radiology
  \textbf{152}(2),  497--501 (1984)

\bibitem{hadlock1985estimation}
Hadlock, F.P., Harrist, R., Sharman, R.S., Deter, R.L., Park, S.K.: Estimation
  of fetal weight with the use of head, body, and femur measurements—a
  prospective study. American journal of obstetrics and gynecology
  \textbf{151}(3),  333--337 (1985)

\bibitem{harrison2017progressive}
Harrison, A.P., Xu, Z., George, K., Lu, L., Summers, R.M., Mollura, D.J.:
  Progressive and multi-path holistically nested neural networks for
  pathological lung segmentation from ct images. In: International conference
  on medical image computing and computer-assisted intervention. pp. 621--629.
  Springer (2017)

\bibitem{van2019automated}
van~den Heuvel, T.L., Petros, H., Santini, S., de~Korte, C.L., van Ginneken,
  B.: Automated fetal head detection and circumference estimation from
  free-hand ultrasound sweeps using deep learning in resource-limited
  countries. Ultrasound in medicine \& biology  \textbf{45}(3),  773--785
  (2019)

\bibitem{jang2017automatic}
Jang, J., Park, Y., Kim, B., Lee, S.M., Kwon, J.Y., Seo, J.K.: Automatic
  estimation of fetal abdominal circumference from ultrasound images. IEEE
  journal of biomedical and health informatics  \textbf{22}(5),  1512--1520
  (2017)

\bibitem{kim2018machine}
Kim, B., Kim, K.C., Park, Y., Kwon, J.Y., Jang, J., Seo, J.K.:
  Machine-learning-based automatic identification of fetal abdominal
  circumference from ultrasound images. Physiological measurement
  \textbf{39}(10),  105007 (2018)

\bibitem{liu2020automated}
Liu, P., Zhao, H., Li, P., Cao, F.: Automated classification and measurement of
  fetal ultrasound images with attention feature pyramid network. In: Second
  Target Recognition and Artificial Intelligence Summit Forum. vol. 11427, p.
  114272R. International Society for Optics and Photonics (2020)

\bibitem{liu2019deep}
Liu, S., Wang, Y., Yang, X., Lei, B., Liu, L., Li, S.X., Ni, D., Wang, T.: Deep
  learning in medical ultrasound analysis: a review. Engineering
  \textbf{5}(2),  261--275 (2019)

\bibitem{long2015fully}
Long, J., Shelhamer, E., Darrell, T.: Fully convolutional networks for semantic
  segmentation. In: Proceedings of the IEEE conference on computer vision and
  pattern recognition. pp. 3431--3440 (2015)

\bibitem{mehta2018net}
Mehta, S., Mercan, E., Bartlett, J., Weaver, D., Elmore, J.G., Shapiro, L.:
  Y-net: joint segmentation and classification for diagnosis of breast biopsy
  images. In: International Conference on Medical Image Computing and
  Computer-Assisted Intervention. pp. 893--901. Springer (2018)

\bibitem{oktay2018attention}
Oktay, O., Schlemper, J., Folgoc, L.L., Lee, M., Heinrich, M., Misawa, K.,
  Mori, K., McDonagh, S., Hammerla, N.Y., Kainz, B., et~al.: Attention u-net:
  Learning where to look for the pancreas. arXiv preprint arXiv:1804.03999
  (2018)

\bibitem{prieto2021automated}
Prieto, J.C., Shah, H., Rosenbaum, A.J., Jiang, X., Musonda, P., Price, J.T.,
  Stringer, E.M., Vwalika, B., Stamilio, D.M., Stringer, J.S.: An automated
  framework for image classification and segmentation of fetal ultrasound
  images for gestational age estimation. In: Medical Imaging 2021: Image
  Processing. vol. 11596, p. 115961N. International Society for Optics and
  Photonics (2021)

\bibitem{ramer1972iterative}
Ramer, U.: An iterative procedure for the polygonal approximation of plane
  curves. Computer graphics and image processing  \textbf{1}(3),  244--256
  (1972)

\bibitem{ravishankar2016hybrid}
Ravishankar, H., Prabhu, S.M., Vaidya, V., Singhal, N.: Hybrid approach for
  automatic segmentation of fetal abdomen from ultrasound images using deep
  learning. In: 2016 IEEE 13th International Symposium on Biomedical Imaging
  (ISBI). pp. 779--782. IEEE (2016)

\bibitem{ronneberger2015u}
Ronneberger, O., Fischer, P., Brox, T.: U-net: Convolutional networks for
  biomedical image segmentation. In: International Conference on Medical image
  computing and computer-assisted intervention. pp. 234--241. Springer (2015)

\bibitem{shah2015perceived}
Shah, S., Bellows, B.A., Adedipe, A.A., Totten, J.E., Backlund, B.H., Sajed,
  D.: Perceived barriers in the use of ultrasound in developing countries.
  Critical ultrasound journal  \textbf{7}(1), ~1--5 (2015)

\bibitem{sinclair2018human}
Sinclair, M., Baumgartner, C.F., Matthew, J., Bai, W., Martinez, J.C., Li, Y.,
  Smith, S., Knight, C.L., Kainz, B., Hajnal, J., et~al.: Human-level
  performance on automatic head biometrics in fetal ultrasound using fully
  convolutional neural networks. In: 2018 40th Annual International Conference
  of the IEEE Engineering in Medicine and Biology Society (EMBC). pp. 714--717.
  IEEE (2018)

\bibitem{sinclair2018cascaded}
Sinclair, M.D., Martinez, J.C., Skelton, E., Li, Y., Baumgartner, C.F., Bai,
  W., Matthew, J., Knight, C.L., Smith, S., Hajnal, J., et~al.: Cascaded
  transforming multi-task networks for abdominal biometric estimation from
  ultrasound  (2018)

\bibitem{wang2018simultaneous}
Wang, P., Patel, V.M., Hacihaliloglu, I.: Simultaneous segmentation and
  classification of bone surfaces from ultrasound using a multi-feature guided
  cnn. In: International conference on medical image computing and
  computer-assisted intervention. pp. 134--142. Springer (2018)

\bibitem{wu2017cascaded}
Wu, L., Xin, Y., Li, S., Wang, T., Heng, P.A., Ni, D.: Cascaded fully
  convolutional networks for automatic prenatal ultrasound image segmentation.
  In: 2017 IEEE 14th international symposium on biomedical imaging (ISBI 2017).
  pp. 663--666. IEEE (2017)

\bibitem{zeng2021fetal}
Zeng, Y., Tsui, P.H., Wu, W., Zhou, Z., Wu, S.: Fetal ultrasound image
  segmentation for automatic head circumference biometry using deeply
  supervised attention-gated v-net. Journal of Digital Imaging pp. 1--15 (2021)

\end{thebibliography}
%




\end{document}